\documentclass{iau}
\usepackage{graphicx}
\usepackage{xcolor}
\usepackage[utf8]{inputenc}
\usepackage[english]{babel}
\usepackage{amsmath}
\usepackage{amsfonts}
\usepackage{amssymb}
\usepackage{subcaption}
\usepackage{hyperref}
\usepackage{wrapfig}

\def\ltsima{$\; \buildrel < \over \sim \;$}
\def\simlt{\lower.5ex\hbox{\ltsima}}
\def\gtsima{$\; \buildrel > \over \sim \;$}
\def\simgt{\lower.5ex\hbox{\gtsima}}

\def\ergs{{erg s$^{-1}$}}
\def\cm2{{cm$^{-2}$}}

\def\p1{{Paper I}}

\def\chandra~{{\em Chandra}}

\def\chandra{{\em Chandra}}

\def\f14{{10$^{-14}$}}
\def\f13{{10$^{-13}$}}
\def\f12{{10$^{-12}$}}
\def\f11{{10$^{-11}$}}
\def\e22{{10$^{22}$}}

\def\l58{{$L_{5.8 \mu m}$}}

\def\msun{{$M_\odot$}}

\def\mbh{{$M_{\rm BH}$}}
\def\lbol{{$L_{\rm Bol}$}}

\def\ledd{{$\lambda_{\rm Edd}$}}

\title[Radio WISSH] 
{Radio WISSH: tuning on the most luminous quasars in the Universe}
\author[Gabriele Bruni]   
{Gabriele Bruni$^1$,
 Javier Mold\'on$^2$,
 Enrico Piconcelli$^3$,
 Francesca Panessa$^1$,
 Miguel P\'erez-Torres$^2$,
 Manuela Bischetti$^4$,
 Chiara Feruglio$^4$,
 Giustina Vietri$^5$,
 Cristian Vignali$^6$,
 Luca Zappacosta$^3$
 \and Ivano Saccheo$^3$
 }
\affiliation{
$^1$INAF -- Istituto di Astrofica e Planetologia Spaziali\\ via del Fosso del Cavaliere 100, 00133 Roma, Italy \\ email: {\tt gabriele.bruni@inaf.it} 
\\[\affilskip]
$^2$CSIC -- Instituto de Astrofísica de Andalucía\\ Glorieta de la Astronomía s/n, 18008 Granada, Spain 
\\[\affilskip]
$^3$ INAF -- Osservatorio Astronomico di Roma\\ via Frascati 33, 00078 Monte Porzio Catone, Italy
\\[\affilskip]
$^4$ INAF -- Osservatorio Astronomico di Trieste\\ via Tiepolo 11, 34143 Trieste, Italy
\\[\affilskip]
$^5$ INAF -- Istituto di Astrofisica Spaziale e Fisica Cosmica di Milano\\ via
A. Corti 12, 20133 Milano, Italy
\\[\affilskip]
$^6$ Dipartimento di Fisica e Astronomia ``Augusto Righi", Università degli Studi di Bologna\\ via Gobetti 93/2, 40129 Bologna, Italy
}

\pubyear{2008}
\volume{xxx}  
\pagerange{119--126}
\jname{Title of your IAU Symposium}
\editors{A.C. Editor, B.D. Editor \& C.E. Editor, eds.}
\begin{document}


\maketitle

\begin{abstract}
In the past years, the results obtained by the WISSH quasar project provided a novel general picture on the
distinctive multi-band properties of hyper-luminous ($L_{bol}>10^{47}$ erg/s) quasars at high redshift (z$\sim$2-4), unveiling interesting relations among active galactic nuclei, winds and interstellar medium, in these powerful sources at cosmic noon.
Since 2022, we are performing a systematic and statistically-significant VLA study of the radio
properties of WISSH. We carried out high-resolution VLA observations aiming at: 1) identifying young radio source
from the broad-band spectral shape of these objects; 2) sample an unexplored high redshift/high luminosity regime, tracking possible
evolutionary effects on the radio-loud/radio-quiet dichotomy; 3) quantifying orientation effects on the observed winds/outflows properties. 
\end{abstract}

\firstsection 

\section{Introduction}

Hyper-luminous QSOs (HyLQSOs, i.e. $L_{bol}>10^{47}$ erg/s), powered by the most massive, highly-accreting supermassive black holes (SMBHs, i.e. $M_{BH} >$ 10$^{9}$ \msun), are ideal targets to probe the assembly of giant galaxies (and, likely, the cradles of proto-clusters). Following the nowadays consensus view on SMBH-host galaxy co-evolution, the huge amount of energy released by highly-accreting SMBHs in HyLQSOs is able to strongly affect the evolution of the host galaxy by heating and expelling  the interstellar medium (ISM) (the so-called ``AGN feedback" mechanism, see e.g. \cite[Fabian 2012]{2012ARA&A..50..455F}, \cite[Morganti 2017]{2017FrASS...4...42M} for a review).
The systematic study of HyLQSOs has faced significant challenges due to their low number density and low fluxes resulting from their distance. A significant  improvement in our understanding of the properties of the accreting SMBH (\mbh, \ledd), the nuclear thermal and non-thermal emission components, multiphase winds and  multiphase ISM (and their interplay) in  HyLQSOs can be only achieved investigating all these aspects in a large sample of HyLQSOs. Namely, one that is markedly different from traditional observing programs in a specific  frequency band, and focused on sparse sources to study a particular aspect of the HyLQSO phenomenon.
This highlights the  necessity of building large samples of HyLQSOs  with extensive multi-band coverage from radio to X-rays. The additional information coming from the radio band can provide fundamental inputs on the presence of jets, their interplay with winds, and in general on the presence of a possible young radio phase at cosmic noon, and the long-standing questions about the radio-loud/radio-quiet dichotomy across cosmic epochs.

\subsection{Probing the brightest end of the AGN luminosity function with WISSH}

The WISSH quasar project can be regarded as a multi-band effort in  the study of HyLQSOs, as demonstrated  by the number of publications since 2017 dealing with their the central engine, the outflows/feedback and host galaxy properties, e.g. \cite{2017A&A...598A.122B,2017A&A...604A..67D,2017A&A...608A..51M,2018A&A...617A..82B,2018A&A...617A..81V,2019A&A...630A.111B,2020A&A...635A.157T,2020A&A...635L...5Z,2021A&A...645A..33B,2022A&A...668A..87V}. The aim of the WISSH project is to establish a reference sample of HyLQSOs at cosmic noon to investigate their nuclear properties and the AGN feedback mechanism on a sound statistical basis. The sample consists of 85 broad-line Type 1, radio-quiet AGN at $z$ $\sim$ 2--4.5 from SDSS-DR7 and selected by the WISE All Sky Survey with flux $F_{22\mu m}>$ 3 mJy (see \cite[Saccheo et al. 2023]{2023A&A...671A..34S} and references therein). Accordingly, the WISSH quasars result to be among the most luminous AGN known in the Universe with \lbol\ $>$ 2 $\times$ 10$^{47}$ \ergs. In this proceeding, we briefly summarize the first results of our radio campaign on WISSH QSOs, aiming at characterizing the radio emission in objects at cosmic noon.


\section{A radio characterization of the WISSH sample}

During 2022, we realized a deep, high-resolution VLA survey of the WISSH sample in A configuration in the 2-8 GHz range. We covered $\sim$90\% of the sample at 2-4 GHz, while $\sim$75\% at 4-8 GHz, probing physical scales between 2 and 5 kpc. Our strategy was to reach a sensitivity threshold of $\sim$50 $\mu$Jy, well below past or current radio surveys. Indeed, a first radio approach to the WISSH sample carried out by \cite{2019A&A...630A.111B} showed that, cross-correlating with the FIRST survey at 1.4 GHz (\cite[Becker et al. 1995]{1995ApJ...450..559B}), only $\sim$20\% of the objects shows a detection at a $\sim$500 $\mu$Jy sensitivity threshold, and a compact morphology ($<$40 kpc at the median redshift of the sample). The recent release of the first epoch of the VLASS survey at 3 GHz - at a sensitivity similar to the FIRST one (RMS$\sim$120 $\mu$Jy/beam) - allowed us to confirm this detection rate. The estimated radio loudness ($R=f_{6cm}/f_{4400\AA}$) is lower than 10 for all except two objects, for which R=47 and 290. Given these premises, going deeper in terms of physical scale and sensitivity appeared to be key to unveil the radio properties of the sample. The main goals of our VLA campaign are the following:

\begin{itemize}
    \item {\underline{Quantify the fraction of young radio sources:}} by compiling the $L_{1.4 \rm{GHz}}$ vs linear size (LS) diagram, we will be able to compare with the fraction of young radio sources found for heavily obscured quasars (\cite[Patil et al. 2020]{2020ApJ...896...18P}), investigating how the different evolutionary stages influence the radio phase. The collected VLA data can be complemented with LOFAR measurements from the LoTSS DR2 survey (\cite[Shimwell et al. 2022]{2022A&A...659A...1S}) - where, among the 34 sources in the footprint, 32 were detected - allowing us to extend the frequency coverage down to 0.15 GHz. The overall radio spectrum will allow us to quantify the fraction of peaked sources in the 0.15-12 GHz range, possibly confirming the fraction of young radio sources estimated from the $L_{1.4 \rm{GHz}}$ vs LS diagram. 
    
    \item {\underline{Probe an unexplored high redshift/high luminosity regime of active galaxies:}} at the median redshift of WISSH ($z=3.33$), and with an RMS of $\sim$10 $\mu$Jy/beam at 3 GHz, it is possible to probe radio powers down to $\sim7\times10^{23}$ W/Hz at 3-sigma significance. Thanks to the availability of optical and X-ray (Chandra program ongoing) luminosities, a distribution of radio-loudness can be obtained and compared to those of other surveys in order to test a possible evolution scenario of the radio loudness (\cite[Ballo et al. 2012]{2012A&A...545A..66B}). This can provide clues on long-standing questions about the radio-loud/radio-quiet dichotomy.
    
    \item {\underline{Test orientation effects on the observed winds/outflows properties:}} in quasars, the spectral index of the optically-thin part of the radio jet spectrum can be used as an indicator of the jet orientation (\cite[Orr \& Browne 1982]{1982MNRAS.200.1067O}), suggesting a near-to-polar line of sight for values $>$--0.5, while an equatorial one for values $<$--0.5 ($S_\nu\propto\nu^\alpha$). This information can be compared with the outflows orientation estimates from \cite{2018A&A...617A..81V}, and with the presence of nuclear winds (BAL) from \cite{2019A&A...630A.111B}. 

\end{itemize}

\subsection{First results: detection rates and morphologies}

\begin{figure}
    \centering
    \includegraphics[width=0.5\textwidth]{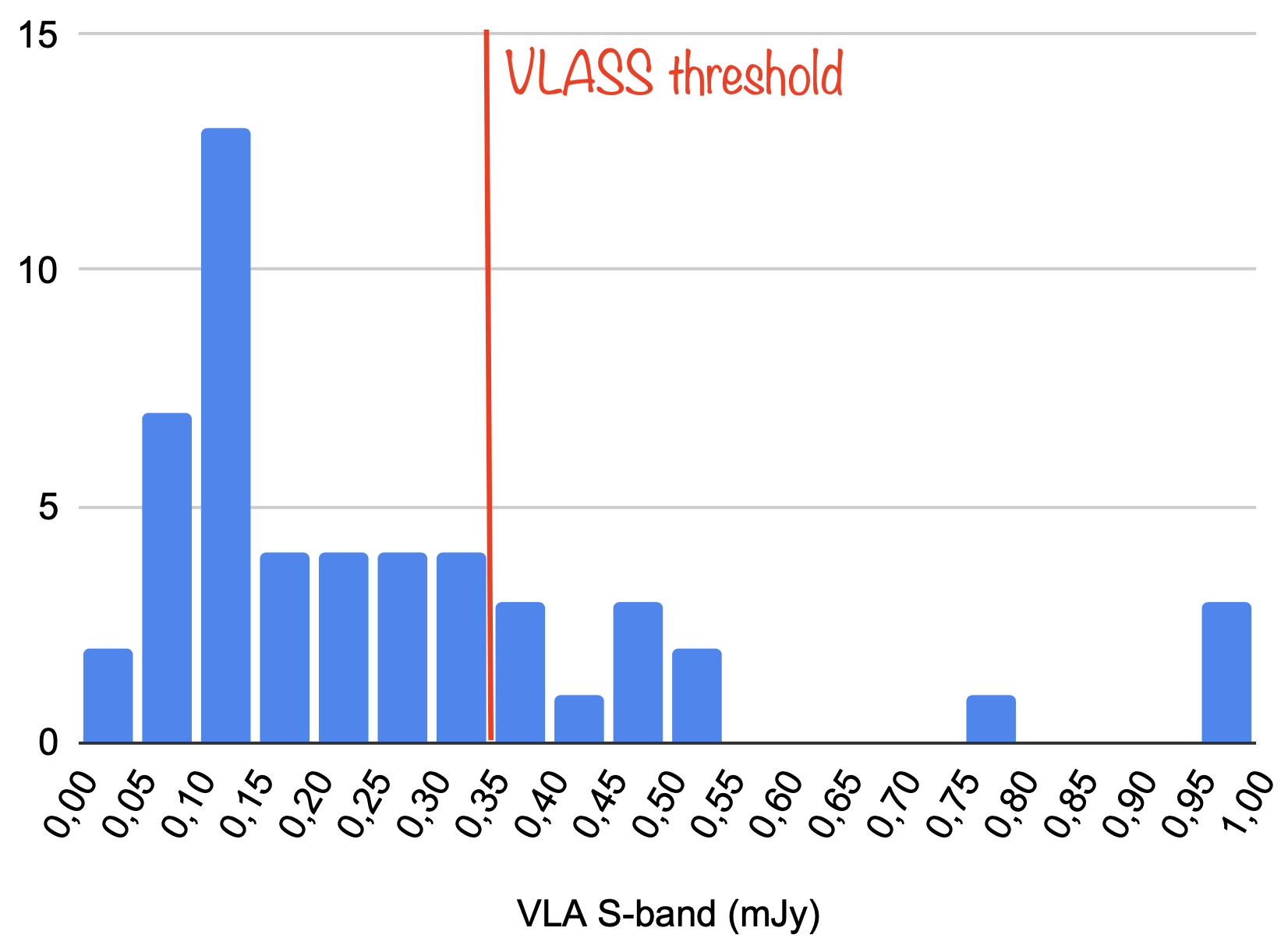}
    \includegraphics[width=0.45\textwidth]{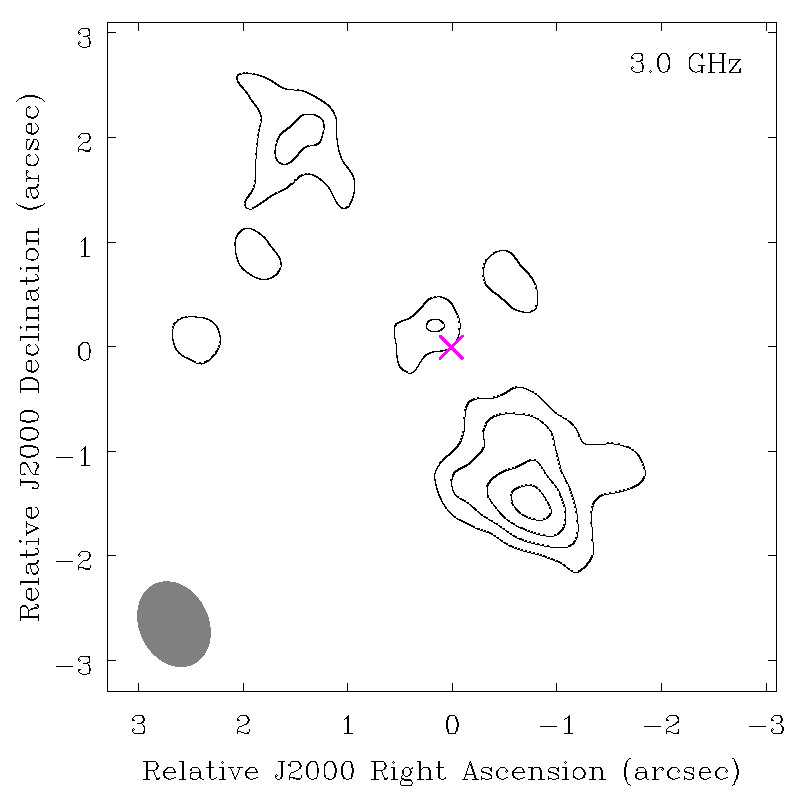}
    \caption{\textit{Left panel} -- Radio flux densities at 3 GHz for the WISSH sample (truncated at 1 mJy). Most of objects lie below the VLASS detection threshold. \textit{Right panel} -- A resolved source from the sample. The cross indicates the optical position of the host, the FWHM is reported in the lower-left corner. The projected linear size of the source is $\sim$40 kpc.}
    \label{fig:wissh}
\end{figure}

At an RMS of $\sim$10 $\mu$Jy/beam, about 80\% of the observed sample was detected at 3 GHz - reaching a redshift of 4.3 - implying an estimated radio power $>10^{23}$ W/Hz (classical threshold between radio-quiet and radio-loud AGN in the local Universe, \cite[Condon 1992]{1992ARA&A..30..575C}). This suggests that, at cosmic noon, most of the hyperluminous AGN like the WISSH ones could host jets. The implication of this result are wide-reaching, from the evolutionary effects on the radio-quiet/radio-loud dichotomy, to the contribution of jets in the QSOs feedback budget at large redshifts, and jets launching at this luminosity regime. The very fact that most of sources lie below the VLASS detection threshold (see Fig. \ref{fig:wissh}, left panel) highlights how important deep observations are to perform population study at this high redshift, allowing to reach the completeness needed to draw conclusions on the radio phase evolution in AGN. 

Three objects showed a resolved morphology at 3 GHz, and more could arise from higher frequency observations. They show a symmetric morphology centered on the optical position of the host, with a projected linear extension of about 30 kpc (see Fig. \ref{fig:wissh}, right panel). This could suggest that these sources are radio galaxies at cosmic noons, but more analysis, including spectral index estimates, is necessary to claim this.  

Observations are still ongoing, and will be concluded at the end of the current VLA semester, completing the multi-frequency radio view on the WISSH sample.


\subsection{The radio phase across cosmic epochs}
Recently, \cite{2020ApJ...896...18P} performed a VLA study of heavily obscured quasars. The sample was selected in the ultra-luminous regime ($L_{bol}\sim10^{11.7-14.2}L_\odot$) at $z\sim0.4-3$, and extremely red in the WISE mid-infrared/optical band, along with a detection of bright, unresolved radio emission from the NVSS. Thanks to high-resolution VLA observations at 10 GHz, they found radio luminosities and linear extents similar to young radio sources (Gigahertz Peaked Spectrum, GPS, and Compact Steep Spectrum, CSS, sources). In a subsequent paper (\cite[Patil et al. 2022]{2022ApJ...934...26P}), they built the radio spectra by adding data from surveys, and confirmed the presence of a high fraction of young radio source.

Although both the \cite{2020ApJ...896...18P} and WISSH quasar samples are selected to allow the direct observation of AGN feedback in action, they are complementary in terms of quasar evolutionary stages. Indeed, according to \cite{2008ApJS..175..390H}, the obscured quasars in \cite{2020ApJ...896...18P} represent the initial heavily dust-enshrouded phase  associated with rapid SMBH growth and star formation triggered by multiple galaxy encounters, while optically-bright objects like WISSH ones are undergoing  the “blow-out” phase, which is characterized by powerful QSO-driven outflows blowing away the nuclear dust cocoon and part of the cold gas reservoir in the host galaxy. 

The same kind of study performed by \cite{2020ApJ...896...18P} and \cite{2022ApJ...934...26P}, once realized on the WISSH sample of hyper-luminous broad-line quasars, will not only provide unprecedented information on the radio phase at cosmic noon, but also shed light on the possible link between the launching mechanism of nuclear winds and radio jets.

\end{document}